\documentclass{LT23auth}
\usepackage{graphicx}
\usepackage{amssymb}

\begin{document}

\begin{frontmatter}

\title{Non-universal critical Casimir force in confined $^4$He
near the superfluid transition}

\author[address1,address2]{X.S. Chen},
\author[address2]{V. Dohm}

\address[address1]{Institute of Theoretical Physics, Chinese
Academy of Sciences, Beijing 100080, China}
\address[address2]{Institut f\"ur Theoretische Physik,
Technische Hochschule Aachen, D-52056 Aachen, Germany}


\begin{abstract}
We present the results of a one-loop calculation of
the effect of a van der Waals type interaction potential $\sim |
{\bf x} |^{-d-\sigma}$ on the critical Casimir force and specific
heat of confined $^4$He near the superfluid transition. We
consider a $^4$He film of thickness $L$. In the region
$L \gtrsim \xi$ (correlation length) we find that the van der
Waals interaction causes a leading non-universal non-scaling contribution of $O
(\xi^2 L^{-d-\sigma})$ to the critical temperature dependence of the
Casimir force above $T_\lambda$ that dominates the universal scaling
contribution $\sim e^{- L/\xi}$ predicted by earlier theories. For the
specific heat we find subleading non-scaling contributions of $O (L^{-1})$
and $O (L^{-d-\sigma})$.
\end{abstract}
%
%
\end{frontmatter}

In the last decade the critical Casimir effect in $^4$He films has
been an interesting topic of theoretical [1,2,3] and
experimental \cite{4} research. On the basis of field-theoretic
calculations for purely short-range interaction it was predicted
that the critical temperature dependence of the Casimir force is
governed by a universal scaling function $X (L / \xi)$ where
$L$ is the film thickness and $\xi = \xi_0 t^{- \nu}$, $t = (T -
T_\lambda) / T_\lambda$, is the bulk correlation length.
The universal behavior for large $L / \xi$ was predicted to be
$\sim e^{- L/\xi}$. Here we investigate the question which
finite-size effect arises from van der Waals type interactions.
They do not change the leading universal {\it bulk} critical
behavior of the free energy and specific heat. The finite-size
effects of these interactions were discussed in Refs.
[1,2,3] where it was assumed that they yield only {\it
non-singular} (temperature-independent) contributions. Here we
shall show that these interactions yield a leading {\it singular}
non-scaling temperature dependence of the Casimir force that is
non-universal and dominates the exponential scaling contribution
above $T_\lambda$ for large $L/\xi$. We also present predictions
for subleading non-scaling contributions to the critical specific
heat of $^4$He films above $T_\lambda$ that may be non-negligible
in the analysis of experimental data.

The present paper is a continuation of recent work where leading
non-scaling cutoff and lattice effects [5,6,7,8] and non-scaling
effects of van der Waals type interactions [8,9,10] have been found
both for the finite-size susceptibility and for the bulk
order-parameter correlation function. These effects were not taken
into account in the previous theory [1,2,3] that was based on the usual
$\varphi^4$ Hamiltonian in $d$ dimensions

%
%

\begin{equation}
H \{\varphi\} = \int d^d x \left[\frac{1}{2} r_0 \varphi^2 +
\frac{1}{2} (\nabla \varphi)^2 + u_0 {(\varphi^2)}^2 \right]
\end{equation}
for the $n$ component field $\varphi ({\bf x})$. Here the
short-range interaction term $(\nabla \varphi)^2$ has the
Fourier representation ${\bf k}^2 \varphi_{\bf k}
\varphi_{-{\bf k}}$. In the following we assume the existence of an
additional subleading long-range interaction term
$b |{\bf k}|^\sigma \varphi_{\bf k} \varphi_{-{\bf k}}$ with $2 < \sigma < 4$
which corresponds to a van der Waals type spatial interaction potential
$U ({\bf x}) \sim |{\bf x}|^{-d-\sigma}$. For the case of $^4$He
$(n = 2)$ we shall consider film geometry with Dirichlet boundary
conditions.

We have performed a one-loop calculation of the critical Casimir
force $F = F_s + F_{ns}$. For the singular part $F_s$ we have found

\begin{equation}
F_{s} (\xi, L, b) = - b L^{- d + 2
- \sigma} B (L / \xi)+  L^{-d} X (L / \xi)
\end{equation}
where the last term $\sim L^{-d}$ is the universal scaling part
due to the short-range interaction. For $\xi \gg L$ this scaling
part is dominant. For large $L/\xi$, however, it becomes
exponentially small, $X (L/\xi) \sim e^{-L/\xi}$ \cite{1}.

The non-universal non-scaling term $\sim b$
due to the van der Waals type interaction reads for $L / \xi
\gtrsim 1$
\begin{equation}
\label{gleichung28} B(L/\xi) = (d - 3 + \sigma) \Psi (L/\xi) -
(L/\xi) \Psi' (L/\xi)\; ,
\end{equation}
\begin{equation}
\label{gleichung24a} \Psi (L/\xi) = \frac{1}{2} (2 \pi)^{\sigma
- 4} \int\limits^\infty_{(L/\xi)^2} dx \left(1 + x
\frac{\partial}{\partial x} \right) \widetilde \Psi (x) \;,
\end{equation}
\begin{eqnarray}
\widetilde \Psi (x) =
\int\limits_0^\infty &dy& \; y^{(2-\sigma)/2}
e^{-x y / 4 \pi^2} \left(\sqrt{\frac{\pi}{y}} \right)^{d-1}
\widetilde W_{1} (y) \times \nonumber \\
&\times& \gamma^* \left(\frac{2-\sigma}{2}\;,
- \frac{x y}{4 \pi^2} \right)
\end{eqnarray}
where $\gamma^* (z, x) = x^{- z} \; \int_0^x \; dt e^{-t} \; t^{z-1}
/\int_0^\infty \; dt e^{- t} \; t^{z-1}$
is the incomplete Gamma function and
\begin{eqnarray}
\widetilde W_1 (y)  = \sqrt{\frac{\pi}{y}}
- \frac{1}{2} \sum^\infty_{n = - \infty} \exp \left(-
\frac{y}{4} n^2 \right) .
\end{eqnarray}
\noindent
The large $L/\xi$ behavior is $\Psi (L/\xi) \sim \xi^2 L^{-2}$.
Thus the leading critical temperature dependence $F_s (\xi, L, b)
\sim b \xi^2 L^{-d-\sigma}$ is nonuniversal for $L \gtrsim \xi$
and dominates the exponentially small scaling part.

Corresponding non-universal non-scaling effects exist for the free
energy \cite{8} which implies a significant restriction for
the range of validity of the universal Privman-Fisher scaling form
\cite{11,12}.

We have also calculated the finite-size effect of the van der
Waals interaction on the specific heat in one-loop order for $L
\gtrsim\xi$ above $T_\lambda$. The result reads
\begin{eqnarray}
C (\xi, L, b) = &C_b& + 2 C_{surface} \left[1 + b
\widetilde A (\sigma, d) \xi^{2-\sigma} \right] L^{-1} \cr
&+& \Delta C + 2 b a_0^2 \widetilde B (\sigma, d) \xi^6 L^{- \sigma - d}
\end{eqnarray}
with $a_0 = (r_0 - r_{0c}) / t$ where $C_b$, $C_{surface}$ and
$\Delta C = L^{\alpha / \nu} g (L / \xi)$ are the known bulk,
surface and finite-size parts of the specific heat due to the
short range interaction \cite{13}. For the relation between $a_0$
and $\xi_0$ see \cite{14}. The amplitudes of the non-universal
non-scaling terms $\sim b$ due to the van der Waals type interaction are
\begin{eqnarray}
\widetilde A (\sigma, d) = - \Gamma \left(\frac{5-d}{2}
\right)^{-1} \int\limits_0^\infty ds \; s^{(3-d-\sigma)/2} e^{-s}
\tilde f_\sigma (s)
\end{eqnarray}
with
\begin{eqnarray}
\tilde f_\sigma (s) &=& \frac{1}{4} \sigma (\sigma-2) \gamma^*
\left(\frac{2-\sigma}{2}, -s \right) + \sigma \gamma^* \left(-
\frac{\sigma}{2}, -s \right) \cr
&+& \gamma^* \left(- 1 - \frac{\sigma}{2}, - s \right)
\end{eqnarray}
and
\begin{eqnarray}
\widetilde B (\sigma, d) = - \frac{\sigma (\sigma-2)}{2^{d+2}
\pi^{d/2}} \; \frac{\Gamma \left(\frac{d + \sigma}{2} \right)} {\Gamma
\left(\frac{4 - \sigma}{2} \right)} \; \zeta (d + \sigma)
\end{eqnarray}
where $\zeta$ is Riemann's zeta function.

The next step of our theory will be to specify the nonuniversal
strength $b$ and the exponent $\sigma$ of the van der Waals type
interaction in $^4$He. Since these parameters are as yet unknown
the range of applicability of the earlier predictions for the
universal finite-size scaling functions of the Casimir force and
of the specific heat above $T_\lambda$ is not yet established.

Finally we note that our one-loop results $\sim b$ were obtained by a
calculation at the Gaussian level (with $u_0 = 0, \; r_{0c} = 0$ and at infinite
cutoff) where part of the non-Gaussian (higher-loop) effects were
taken into account by the replacement $r_0 \to \xi^{-2}$. It remains to
be seen to what extent a full renormalization-group treatment with
$u_0 > 0$ at the two-loop level confirms this procedure.In this sense
our predictions of the temperature dependence
of the terms $\sim b$ in Eqs. (2) and (7) should be considered
only as preliminary. A full renormalization-group treatment will
be given elsewhere.

%
%
\begin{ack}

We acknowledge support by DLR and by NASA under grant numbers
50WM9911 and 1226553.

\end{ack}

%
%

\end{document}